\documentclass[reprint, amsmath,amssymb,prl]{revtex4-2}

\usepackage{graphicx}
\usepackage{dcolumn}
\usepackage{bm}
\usepackage{gensymb}
\usepackage[colorlinks=true,linkcolor=blue, citecolor=blue]{hyperref}
\graphicspath{{./fig_rl/}}
\usepackage{xcolor}
\begin{document}
	
	
	\title{Tailoring photostriction via superlattices engineering}

\author{Carmel Dansou\textsuperscript{1}, Charles Paillard\textsuperscript{1,2}, Laurent Bellaiche\textsuperscript{1,3}}
\affiliation{\textsuperscript{1}Smart Ferroic Materials Center, Institute for Nanoscience \& Engineering and Department of Physics, University of Arkansas, Fayetteville, Arkansas 72701, USA}
\affiliation{\textsuperscript{2}Universit\'{e} Paris-Saclay, CentraleSup\'{e}lec, CNRS, Laboratoire SPMS, 91190 Gif-sur-Yvette, France.}
\affiliation{\textsuperscript{3}Department of Materials Science and Engineering, Tel Aviv University, Ramat Aviv, Tel Aviv 6997801, Israel.}

	\date{\today}
	
	\begin{abstract}
		We report  systematic first-principles investigation of light-induced mechanical deformations in monodomain (PbTiO\ensuremath{_{3}})\ensuremath{_{n}}/(SrTiO\ensuremath{_{3}})\ensuremath{_{n}} superlattices ($n=1-5$). We reveal that photostriction in these heterostructures quantitatively and qualitatively depends on the chemical period $n$.  Specifically, we show that by changing the chemical period, we can induce {\it positive} or {\it negative} photostriction. We also present a simple analytical model to account for the calculated deformations.  Our findings indicate that superlattices architectures may be key to design novel optomechanical applications.
	\end{abstract}
	
	\maketitle

Photostriction is a physical phenomenon in which light induces mechanical strains in a material \cite{ku20}.  It therefore converts light into mechanical energy \cite{ku20,chen202,finkel}. In recent years,  materials such as ferroelectrics \cite{Haleoot2017,Paillard2017,Paillard2016,matz,gao2023}, multiferroics \cite{kund,kundlig,wen2013,schick2014,pail1,gu2023}, organic polymers \cite{naka}, inorganic semiconductors \cite{busch} and halide perovskites \cite{zhou20, paillard2023,bpeng20} were investigated for their photostrictive response. 
Photostrictive properties can be harnessed for practical applications such as optomechanical sensors \cite{datskl}, energy harvesting devices \cite{wang20}, and phostrictive actuators \cite{poos,sun}, among others. 

In the last decade, the field of ferroelectrics has also witnessed the emergence of ferroelectrics/dielectrics (FE/DE) heterostructures. They host exotic properties that make them prime candidates for the next generation of electronic devices \cite{da1,da2,ag1,bq,co,das,yad,abid,ag2,zub1}.  Recent experiments have shown that FE/DE superlattices (SLs) also exhibit lattice deformations upon optical excitation \cite{ahn2017,lee2021}.  Due to the fine,  atomic layer control achieved by nowadays state-of-the-art growth techniques, SLs offer a unique playground to engineer, from the bottom-up, the photostrictive response. 
However, existing experimental reports are limited to as-grown SLs, with no systematic exploration of the parameters involved in their photo-induced deformations. For instance, only light-induced lattice expansion has been reported in the prototype PbTiO\ensuremath{_{3}}/SrTiO\ensuremath{_{3}} (PTO/STO) SLs \cite{dar,ahn2017}. First-principles calculations, as conducted here, allow on the other hand to explore the space of parameters while saving time and avoiding material waste.

In this Letter, we propose a first-principle study of photostrictive behavior of single domain (PbTiO\ensuremath{_{3}})\ensuremath{_{n}}/(SrTiO\ensuremath{_{3}})\ensuremath{_{n}}  (($n|n$) PTO/STO) SLs and show that, by changing the chemical period $n$ of the SLs, the magnitude and even {\it sign} of the photo-induced deformations can be tuned. We also develop a simple phenomenological model that reproduces the observed unusual photostrictive behaviors qualitatively and quantitatively. 

We use the Abinit code \cite{ab} for our density functional theory (DFT) calculations. We employ the constrained occupation number approach developed in Ref.~\cite{paillard2019}, and thus enforce $n_{ph}$ electrons (resp. holes) to stay in the conduction (resp. valence) bands using a Fermi-Dirac distribution with its own quasi-Fermi level $\mu_e$ (resp. $\mu_h$).
 Here, we consider $(n|n)$ SLs with polarization along the [001] direction, generally referred to as $c$ phase in literature \cite{pertsev}. We vary the  periodicity $n$ from $1$ to $5$ (and thus the chemical period from $2$ to $10$ perovskite cells).  We thus sample SLs with electrostatic coupling between the PTO and STO layers varying from strong ($n\le3)$ to weak $(n\ge3)$ as defined in Ref.~\cite{zubko20}.  Polarization is induced by displacing Ti atoms in the PTO layers up along the z-axis from their centrosymmetric position at the beginning of our simulations. The technical details and convergence parameters are the same as presented in the companion paper \cite{xxx}.

\begin{figure*}[ht!]
	\centering
	\includegraphics[width=1\linewidth]{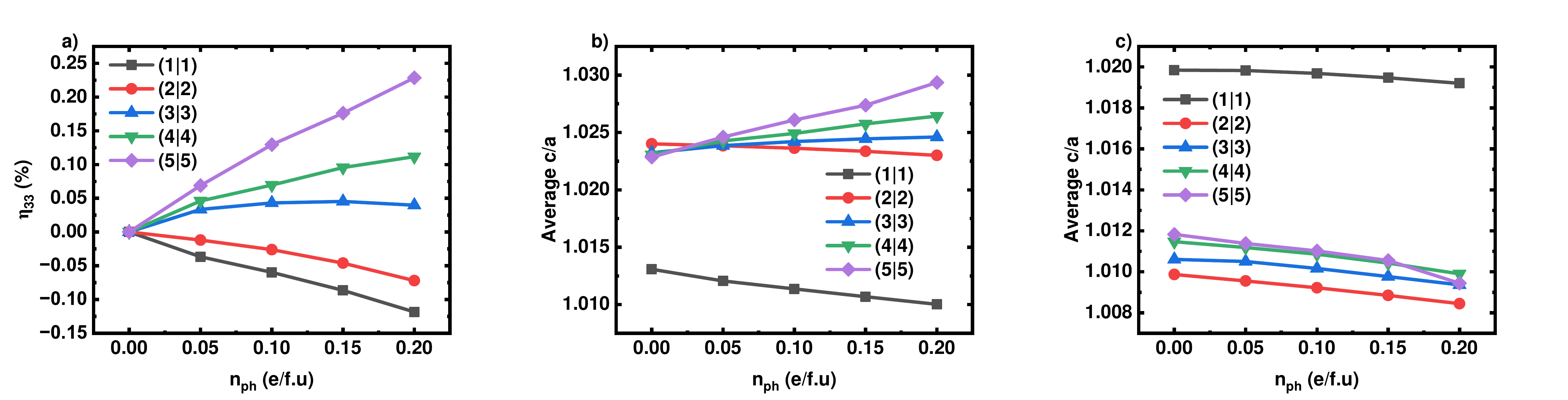}
	\caption{\label{f1} a) Lattice deformation in the $(n|n)$ SLs as function of $n_{ph}$. b) Average c/a of the PTO layer as function of  $n_{ph}$, c)  Average c/a of the STO layer as function of $n_{ph}$.}
\end{figure*}

Let us first look at the 
photo-induced strain $\eta_{33} = (c(n_{ph})-c(0))/c(0)$ (where $c(n_{ph})$ is the out-of-plane lattice constant at the concentration of photo-excited carriers $n_{ph}$) 
as a function of $n$ in the $(n|n)$ SLs. As shown in Fig. \ref{f1}, one can distinguish two behaviors when focusing on the sign of the light-induced strain.
For the ultrathin chemical period $(n=1,2)$ SLs i.e., in the strong electrostatic coupling regime of the PTO and STO layers, light induces a compression, \textit{i.e.} {\it negative} photostriction (see Fig \ref{f1}(a) for different chosen values of $n_{ph}$).  In contrast, photostriction is {\it positive} (\textit{i.e.,} the SLs expands in the out-of-plane direction) for thicker SLs $n\ge 3$ (\textit{i.e.,} in the weak electrostatic coupling regime of the PTO and STO layers). 
To investigate in more details these two regimes,  we examine the light-induced tetragonal distortion in the different layers. Consequently, in Fig \ref{f1}(b) \& c) is shown the average tetragonal distortion in the PTO and STO layers of the SLs as a function of the concentration of photoexcited carriers for all values of $n$.  STO is an incipient ferroelectric material, however, the proximity with PTO induces {in DFT calculations, a polarization in  the STO layer (via an internal polarizing field) and thus a significant tetragonal distortion. For utrathin SLs with $n=1,2$, both the distortion in the PTO and STO layers (black and red curves in Figs \ref{f1} b\&c) decrease under optical excitation leading to the overall observed compression in the out-of-plane lattice constant; a feature reminiscent of the behavior of bulk ferroelectric titanates~\citep{Paillard2017,paillard2019}. On the other hand, in the SLs  $n\ge 3$, upon optical excitation, the distortion in the STO layer decreases whereas the distortion in the PTO layer increases (see blue, green and purple curves in Fig. \ref{f1} b\&c). The positive distortion in the PTO layers dominates and thus the overall expansion is observed. The data depicted in Fig.~\ref{f1} thus suggest that the photostrictive response (negative or positive) of the PTO layer and the SL as a whole is strongly dependent on the strong or weak electrostatic coupling between the layers.

We now examine the evolution of the electrical polarization of the SLs as a function of the concentration of photo-excited carriers. The polarization in the SLs is computed in a layer-by-layer fashion using the following equation:
\begin{equation}\label{eq1}
	P_{i} = \dfrac{e}{\Omega_{i}}\sum_{j\in i} w_{j}\delta_{j} Z^{*}_{j} 
\end{equation}
where $\Omega_{i} $, $e $, and $\delta_{j} $ are the volume of the local unit cell $i$ (local unit cell is used here to refer to an effective five atoms unit cell centered on a Ti atom), the elementary charge and the atomic displacement from a built reference nonpolar SLs (we built this reference structure for each of the SLs) respectively. The weight factor $w_{j}$ tells how many of such local cells share the $j$ ions. Its values are $1/2$, $1$, $1/8$ for the O, Ti and (Pb,Sr) atoms respectively. $Z^{*}_{j} $ is the $z$ component of the Born effective charge of atom $j$, with $j$ running through all the atoms in the local cell $i$. The Born effective charges $Z^{*}_{j} $ are those of the atoms in dark conditions and are taken from Ref.~\cite{zho}. We do not expect a significant alteration of $Z_j^{*}$ under the concentration of photo-excited carriers investigated here, and thus do not employ the more refined approach devised in Ref.~\cite{dreyer}. The total polarization in the SLs is the average of the polarization in PTO and STO layers.
\begin{figure*}[ht!]
	\centering
	\includegraphics[width=1\linewidth]{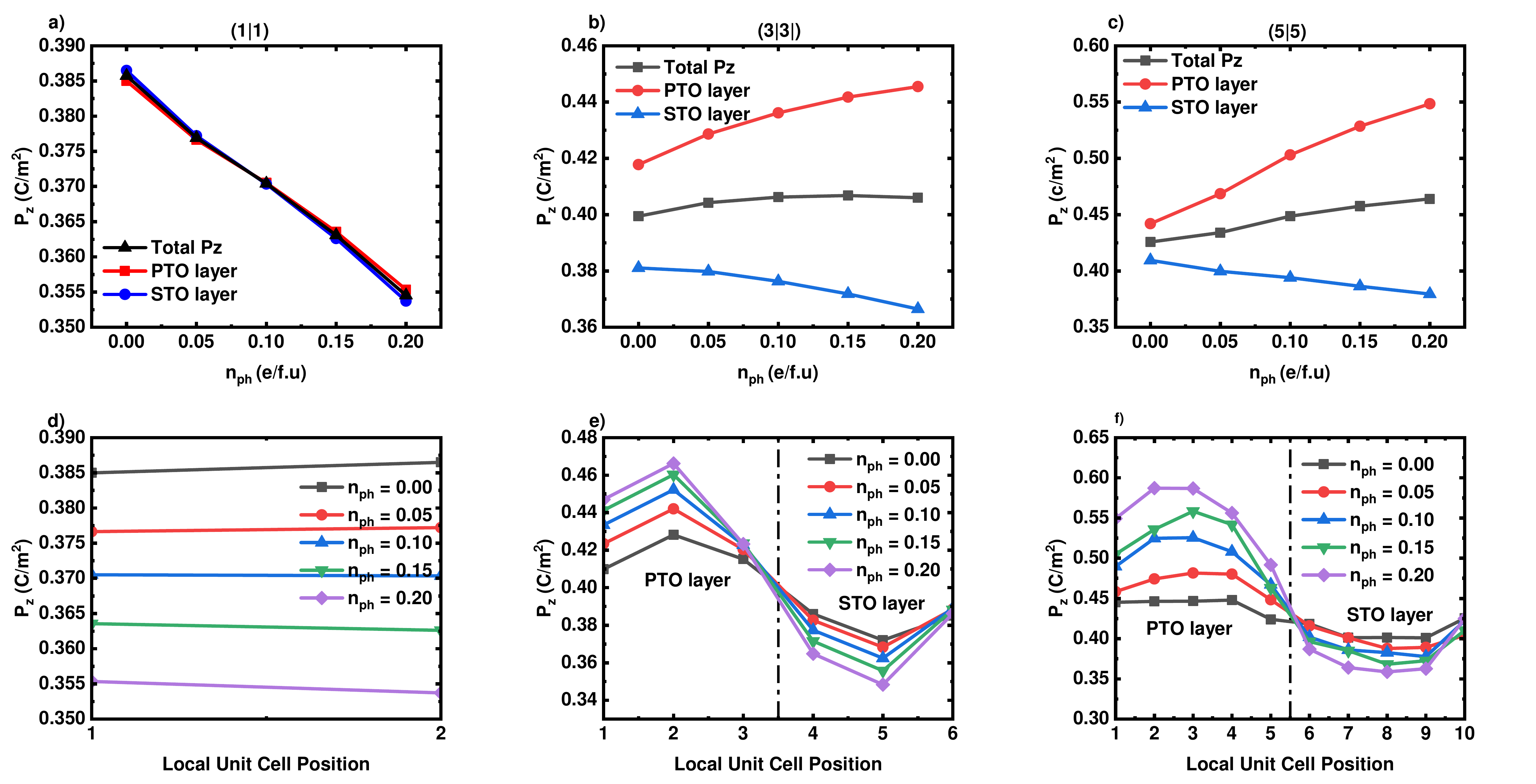}
	\caption{\label{f2} Top panel: polarization as a function of $n_{ph}$ in the $(n|n)$SLs. a)-c) $(1|1)$, $(3|3)$, and $(5|5)$ SLs respectively. Bottom panel: projection to five-atom unit cell of the polarization in the d)-f) $(1|1)$, $(3|3)$, and $(5|5)$ SLs respectively. The vertical dashed lines in e) and f) are to separate local cells corresponding to PTO and STO layer in each SLs.}
\end{figure*}
As shown on Fig \ref{f2} for selected SLs (See Supplementary Material \cite{sprl}} for the other SLs), optical excitation reduces the total polarization in ultra-thin ($n|n$) SLs ( $n=1$ and $n=2$). Additionally, polarization in both the STO and PTO layers decreases upon optical excitation. This is consistent with the decrease of the lattice distortion shown on Fig \ref{f1} (b\& c) for the $(1|1)$ and $(2|2)$  SLs. When the SLs layers are thick enough, that is for $n\ge 3$ , the overall polarization is enhanced by optical excitation as shown by the black curves on Figs \ref{f2}  (b\&c). Additionally, the polarization in PTO layer increases (see red curves on Fig \ref{f2} b \& c) whereas the induced polarization in the STO layer is decreased (see blue curves on Fig \ref{f2} b \& c) -- implying that STO becomes less polar under light.

To gain further insight on the role of the free charges in the above observed changes in the lattice constant and polarization inside the SLs, we computed the macro-average \cite{j7} excited charges density and the electrostatic potential inside the SLs under illumination for the considered ($n|n$) SLs. 
\begin{figure}[ht!]
	\centering
	\includegraphics[width=1\linewidth]{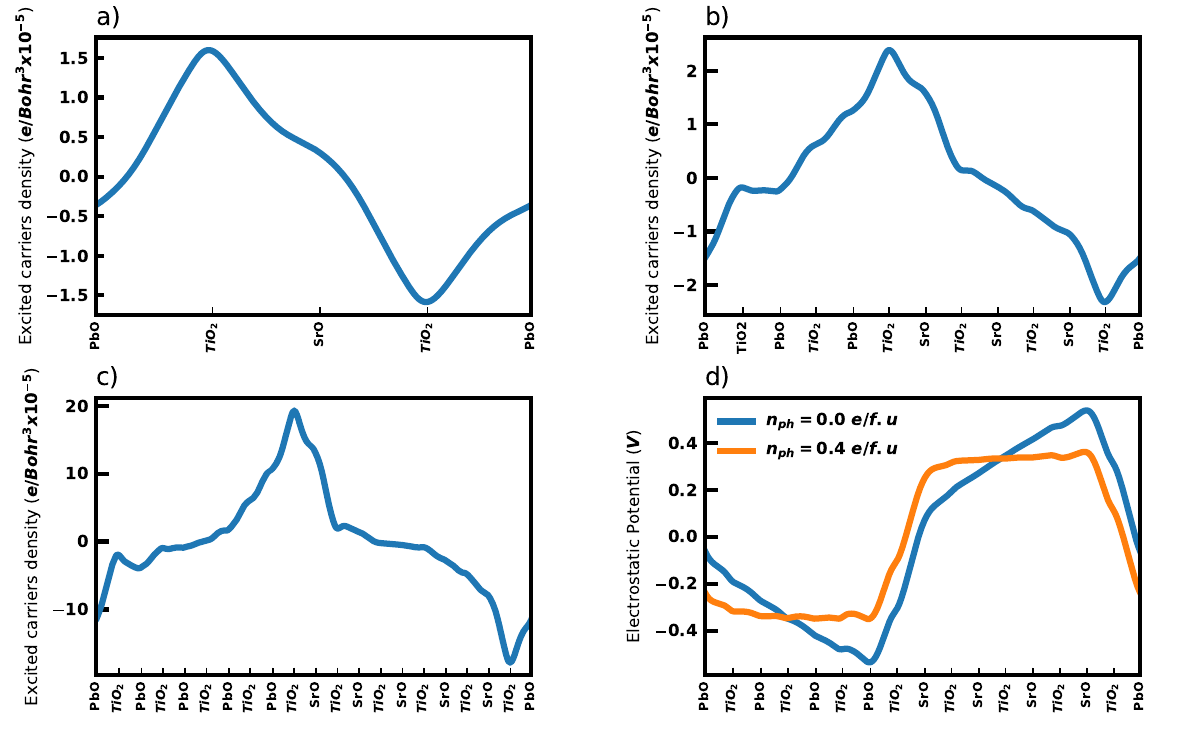}
	\caption{\label{f3} Excited carriers density for a) $(1|1)$ SLs, b) $(3|3)$ SLs, c) $(5|5)$ SLs, computed at $n_{ph} = 0.4$ e/f.u. (d) Electrostatic potential in dark (blue) and under illumination (orange) for the $(4|4)$ SLs. Plots in a)- c) are the difference between the macro-average density under illumination and in dark.}
\end{figure}
As shown in Fig \ref{f3} (a)-(b), in the ultrathin $(n|n)$ SLs ($n=1$ or $n=2$), the photo-excited carriers  are more delocalized through the SLs than for $n \ge 3$. When the thickness of the films is increased ($n$ is equal or greater than 3), the free charges become strongly localized on TiO\textsubscript{2} planes forming the interface, as shown in Fig. \ref{f3} (a) to (c) for the $(1|1)$, $(3|3)$ and $(5|5)$ SLs respectively (see the Supplemental Materials \cite{sprl} for $(2|2)$ and $(4|4)$ SLs). The spacial distribution of the free charges in the $(n|n)$ SLs provide us with the following pictures: the ultrathin SLs, where there is a strong electrostratic coupling between PTO and STO layers, behave like bulk system~\cite{pail1,Paillard2017,paillard2019} where photo-excited carriers are spread out within the SL and lead to an overall decrease of polarization and thus out-of-plane lattice compression in both PTO and STO layers.
For thick enough SLs, the free charges strongly localize at the interfaces. Electrons and holes populate given interfaces and produce fields that oppose the existing field in the STO and PTO layers within the SLs. 
The light-induced field in STO thus opposes the polarizing field and leads to the decrease in the polarization in STO and to the observed compression of this specific layer.  On the other hand, the light-induced field in PTO opposes the depolarizing field there too, which then results in an increase of the electrical polarization, therefore generating the observed expansion of the PTO layer (see the Supplemental Materials \cite{sprl} for a schematic). We further elucidate this by mapping out the electrostatic potential in the SLs before and after illumination (see Fig. \ref{f3}d for the $(4|4)$ SLs, the other are shown in the supplemental materials). In dark, the electrostatic potential has a non-zero slope in both the STO and PTO layers (see blue curve in Figure~\ref{f3}d). This suggests the presence of a finite field as postulated above. Upon illumination, the electrostatic potential flattens leading to vanishingly small electric fields within layers of the SLs (see orange curve in Fig \ref{f3}d in the case of the $(4|4)$ SLs).  Essentially, in thick SLs, photo-excitation provides enough free carriers to screen the polarization bound charges and quench the resulting local electrostatic fields in each layer.

\begin{figure*}[ht!]
	\centering
	\includegraphics[width=1\linewidth]{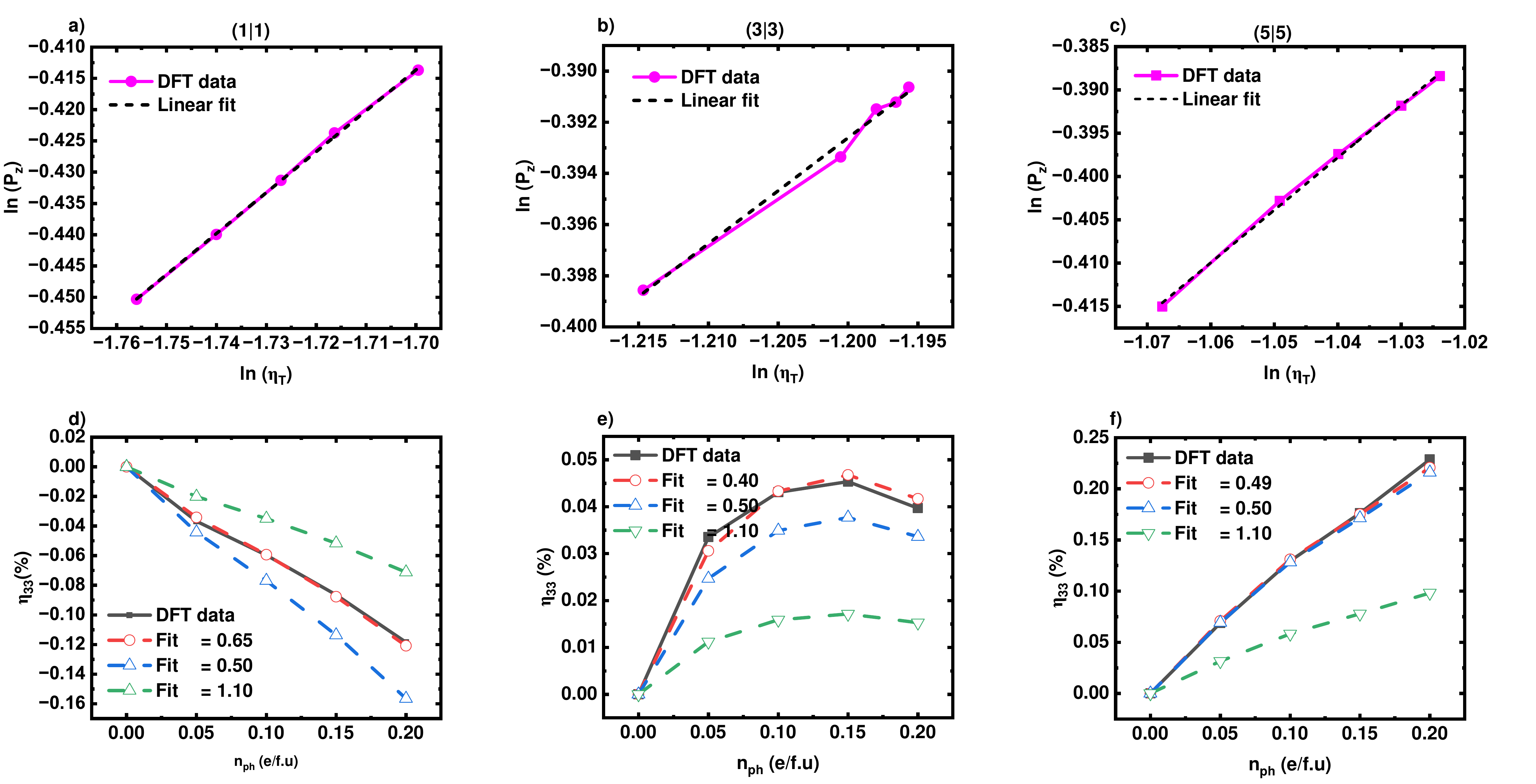}
	\caption{\label{f4} Top panel a) to c) plot of $\ln(P)~ vs ~\ln(\eta_{T})$ for the $(1|1)$, $(3|3)$ and $(5|5)$ SLs respectively. Bottom panel d) to f) plot of the DFT data and fit using eq \ref{eq3} for different values of $\alpha$ for the $(1|1)$, $(3|3)$ and $(5|5)$ SLs respectively.}
\end{figure*}

Previous studies based on Landau-Guinzburg-Devonshire (LGD) theory have shown that, when the direction of the polarization is locked along the $z$ axis (the argument is also true for the other axes), the tetragonality of the SLs can serve as a probe of polarization of the SLs \cite{da1,da2}. It was established that the polarization is proportional to an exponent of the tetragonality $\eta_{T} = (c-c_{p})/a$ with $c$ and $c_{p}$ being the out-of-plane lattice constant of the polar SLs and the corresponding paraelectric SLs, and $a$ the in-plane lattice parameter of the SLs. Earlier works on single domain PTO/STO SLs in dark have shown that $P_{z} \propto \eta_{T}^{1/2}$ \cite{da2}, however, in a recent work, it was found that $\alpha$ differs from $1/2$ when free charges screening effect is considered. The study found that the values of $\alpha$ depends on the SLs, its structure and on the pump used for the excitation. Typically, in presence of free charge, it was found that $P_{z} \propto \eta_{T}^{1.1}$ \cite{lee2021} for a polydomain (PbTiO\ensuremath{_{3}})\ensuremath{_{8}}/(SrTiO\ensuremath{_{3}})\ensuremath{_{3}} SLs. In view of these previous works, we plot $\ln(P_{z}(n_{ph})) $ as function $\ln(\eta_{T} (n_{ph}))$ for the considered SLs. As shown on the top panel of Fig \ref{f4} (shown for $(1|1)$, $(3|3)$ and $(5|5)$ SLs), there is a linear relation between $\ln(P_{z}(n_{ph})) $ and $\ln(\eta_{T} (n_{ph}))$ -- implying that $P_{z} \propto \eta_{T}^{\alpha} = \left[ (c-c_{p})/a \right]^{\alpha}$. We obtained $\alpha$ (0.65, 0.56, 0.40, 0.62 and 0.49 for the $(1|1)$, $(2|2)$, $(3|3)$, $(4|4)$ and $(5|5)$ respectively) by taking the slope of a linear fit into $\ln(P_{z}(n_{ph}))~ vs ~\ln(\eta_{T} (n_{ph}))$ for each SLs.
\begin{equation}\label{eq1}
	 P_{z} \propto \eta_{T}^{\alpha} = \left[ (c-c_{p})/a \right]^{\alpha}. 
\end{equation}
Taking the logarithmic derivative of Eq \ref{eq1}, gives the following:
\begin{equation}\label{eq2}
	\dfrac{dP_{z}}{P_{z}} = \alpha \dfrac{d c}{(c-c_{p})}.
\end{equation}
From Eq \ref{eq2}, it can be readily established that the strain $dc/c$ in the out-of-plane lattice constant is given by
\begin{equation}\label{eq3}
	dc/c =\eta_{33}= \dfrac{\eta_{d}}{\alpha} \dfrac{dP_{z}}{P_{z}}.
\end{equation}
where we define $\eta_{d} = (c-c_{p})/c $. The scaling relation in Eq. \ref{eq1} provides us with a simple analytical expression of the light induced strain $\eta_{33}$ as function of the light-induced relative change in the polarization. The estimated strains using Eq. \ref{eq3}, along with the DFT data, are shown on the right panel in Fig \ref{f4} for each of the SLs. It is in excellent agreement with the DFT data. Also we plot the estimated strain using reported values of $\alpha$ from Refs. \cite{da2,lee2021} for comparison.

Although not the focus of this study, we would like to comment on the obtained values of $\alpha$. As found in Ref. \cite{lee2021}, it would be expected that $\alpha$ depends on the SLs. Also one would expect the values of $\alpha$ to deviate from $1/2$ as we find in our study. However, with the low excited carriers concentration considered in this study and the simple structure of the SLs, we do not expect a drastic deviation from $1/2$, consistent with the range of values $0.4-0.65$ found for the considered $(n|n)$ corresponding to a maximum deviation of $0.15$ from $0.5$. As discussed in Ref. \cite{lee2021}, the physical significance of $\alpha$ is unclear to us. It would interesting to investigate in more details the effect of free charge on $\alpha$ and its behavior as function of the SLs.
  
In summary, we study, using first-principle calculations, light-induced deformation in single domain $(n|n)$ PTO/STO SLs $(n=1-5)$. We show that, for very thin SLs ($n\leq 2$), free carriers are delocalized and screen existing dipoles, which induces contractions in the out-of-plane lattice of the SLs. On the other hand, when the film is thick enough ($n \geq 3$), the free charges are localized in the interfaces regions. They produce a field that opposes existing depolarizing fields in the SLs and thereby induces a contraction in the STO layers and oppositely an expansion in the PTO layers and leads to an overall expansion of the SLs. These show that the photostrictive behavior of PTO/STO SLs can be tuned through its thickness. Our findings show a promising route to controlling and tuning the light-induced strain in PTO/STO SLs by simply changing its thickness making them appealing for potential applications using light as a handle.

\begin{acknowledgements}
	The work is supported by the ARO Grants No. W911NF-21-2-0162 (ETHOS) and W911NF-21-1-0113, and the Vannevar Bush Faculty Fellowship (VBFF) grant no. N00014-20-1-2834 from the Department of Defense.  C.P.  thanks the support from a public grant overseen by the French Agence Nationale de la Recherche under grant agreement no. ANR-21-CE24-0032. We also acknowledge the computational support from the Arkansas High Performance Computing Center (AHPCC) for computational resources.	
\end{acknowledgements}

	\bibliography{PRL}
	
\pagebreak
\widetext
\begin{center}
\textbf{\large Supplemental Materials for ``Tailoring photostriction via superlattices engineering"}
\end{center}
\section{Polarization in the $(2|2)$ and $(4|4)$ SLs as function of $n_{ph}$.}
\begin{figure}[ht!]
	\centering
	\includegraphics[width=0.85\linewidth]{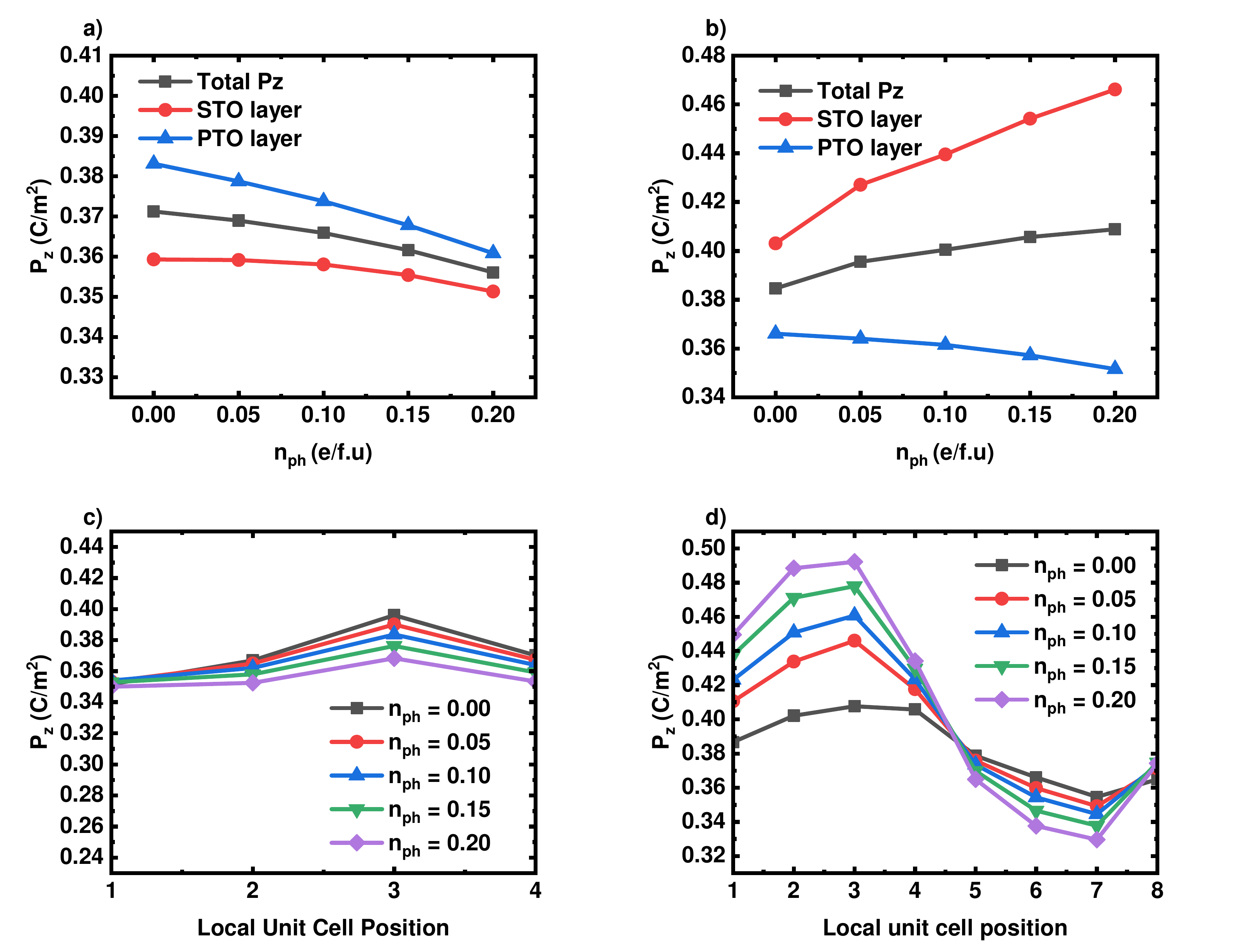}
	\caption{Polarization as a function of $n_{ph}$ in the $(n|n)$SLs. a) $(2|2)$, b) $(4|4)$ SLs respectively. Projection to five-atom unit cell of the polarization in the c) $(2|2)$, d) $(4|4)$ SLs respectively.}
\end{figure}
\newpage
\section{Schematic diagram of the action of light in the $(n|n)$ SLs}
\begin{figure}[ht!]
	\centering
	\includegraphics[width=0.85\linewidth]{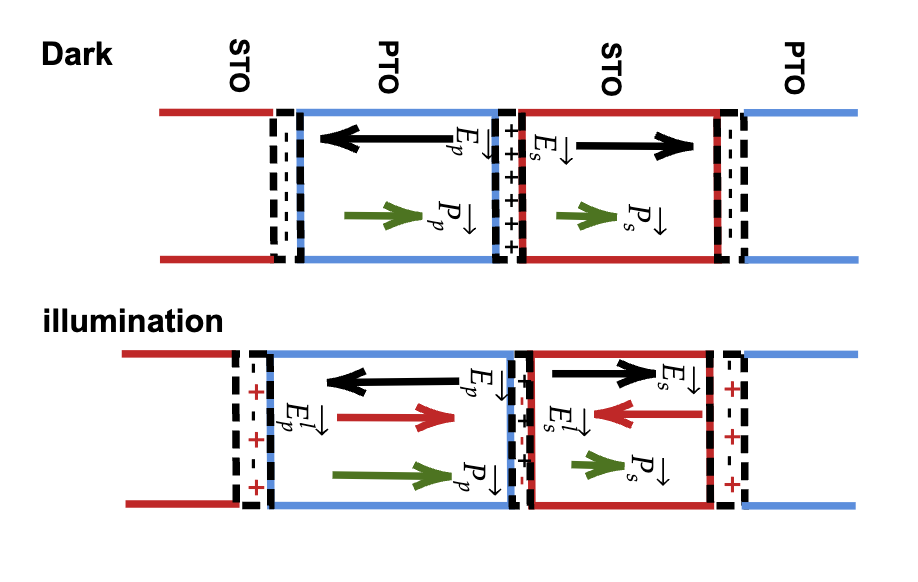}
	\caption{Schematic diagram of the action of light in the $(n|n)$ SLs. ($\vec{E}_{p},~\vec{P}_{p}$) and ($\vec{E}_{s},~\vec{P}_{s}$) are the coupled field and polarization in PTO and STO layer in dark conditions, respectively. $\vec{E}_{p}^{l},~\vec{E}_{s}^{l}$ are the light-induced field in PTO and STO layer, respectively.}
\end{figure}

\section{Photoexcited charges density in $(2|2)$ and $(4|4)$ SLs computed at $n_{ph}=0.4$}
\begin{figure}[ht!]
	\centering
	\includegraphics[width=1\linewidth]{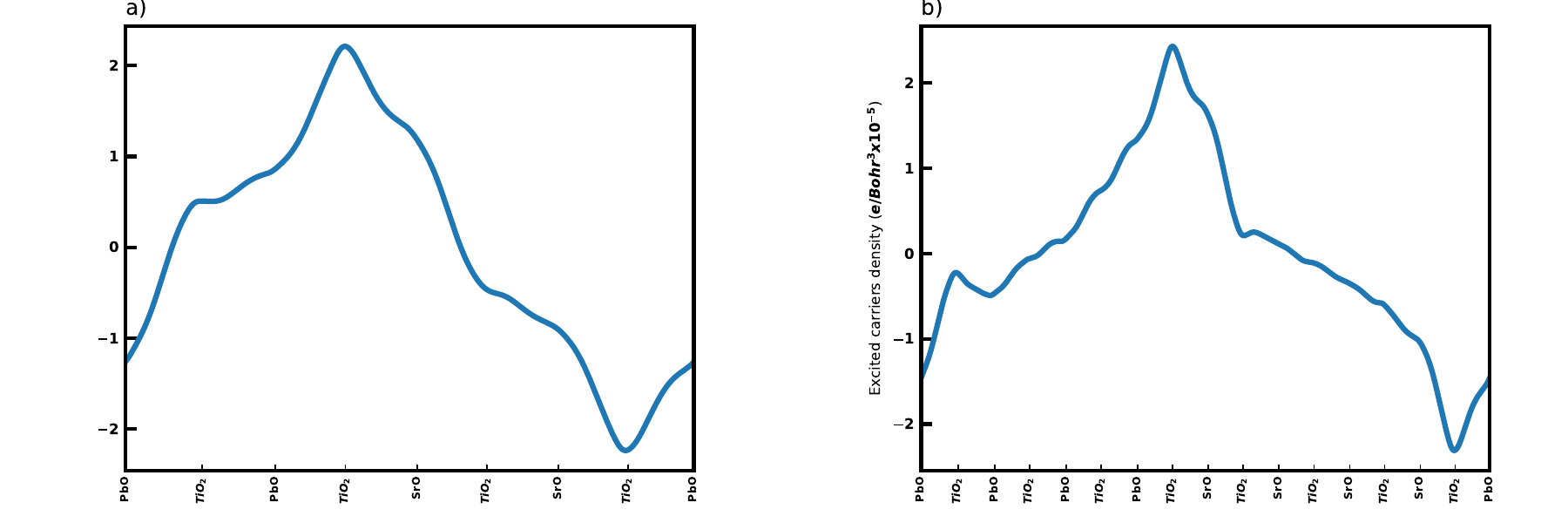}
	\caption{Excited carriers density for a) $(2|2)$ SLs, b) $(4|4)$  SLs, computed at $n_{ph} = 0.4$ e/f.u.}
\end{figure}

\newpage
\section{Values of $\alpha$ as function of $n$ in the $(n|n)$ SLs}
\begin{figure}[ht!]
	\centering
	\includegraphics[width=0.5\linewidth]{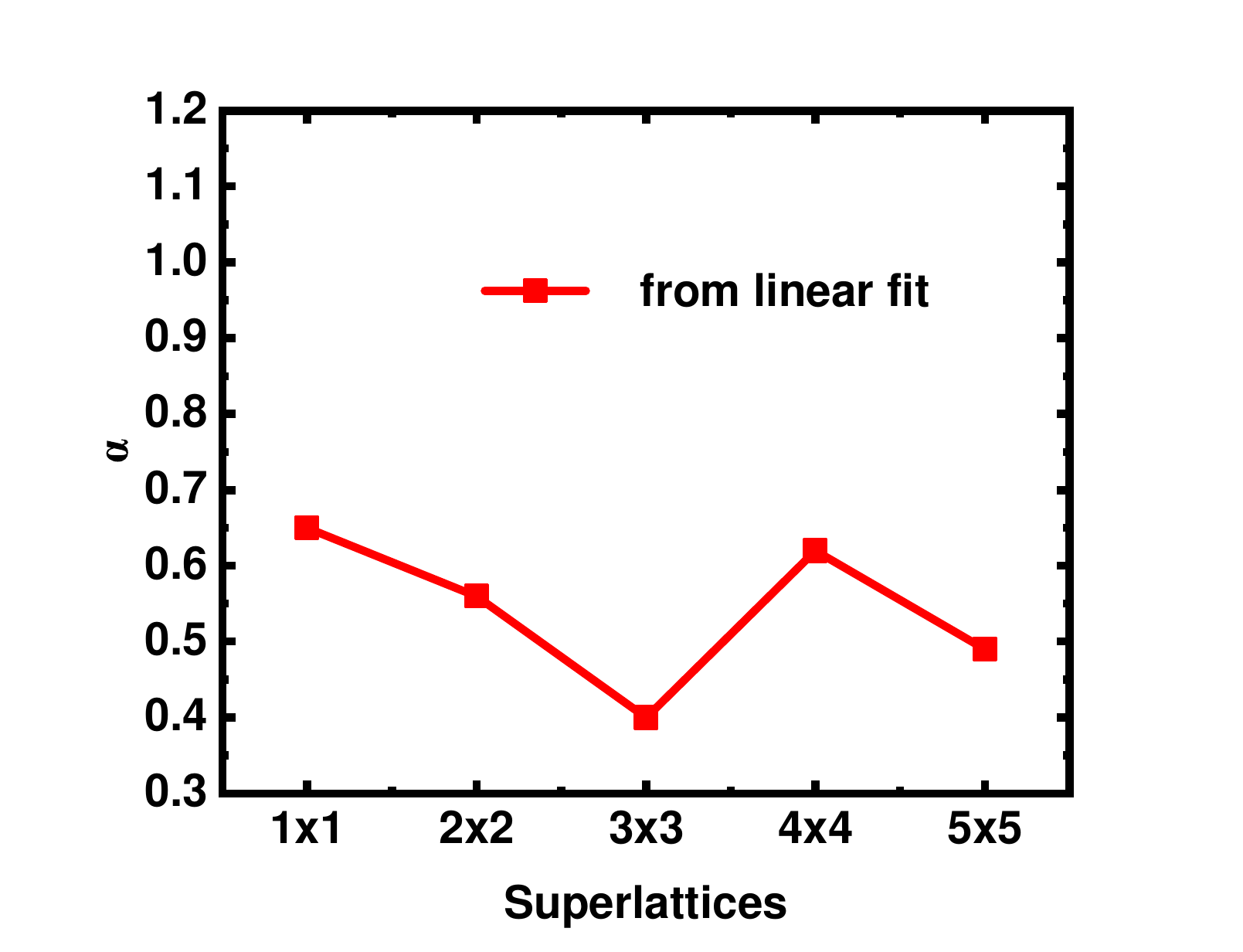}
	\caption{Values of $\alpha$ as function of the $(n|n)$ SLs}
\end{figure}

\section{Model results for the $(2|2)$ and $(4|4)$ SLs}
\begin{figure}[ht!]
	\centering
	\includegraphics[width=.75\linewidth]{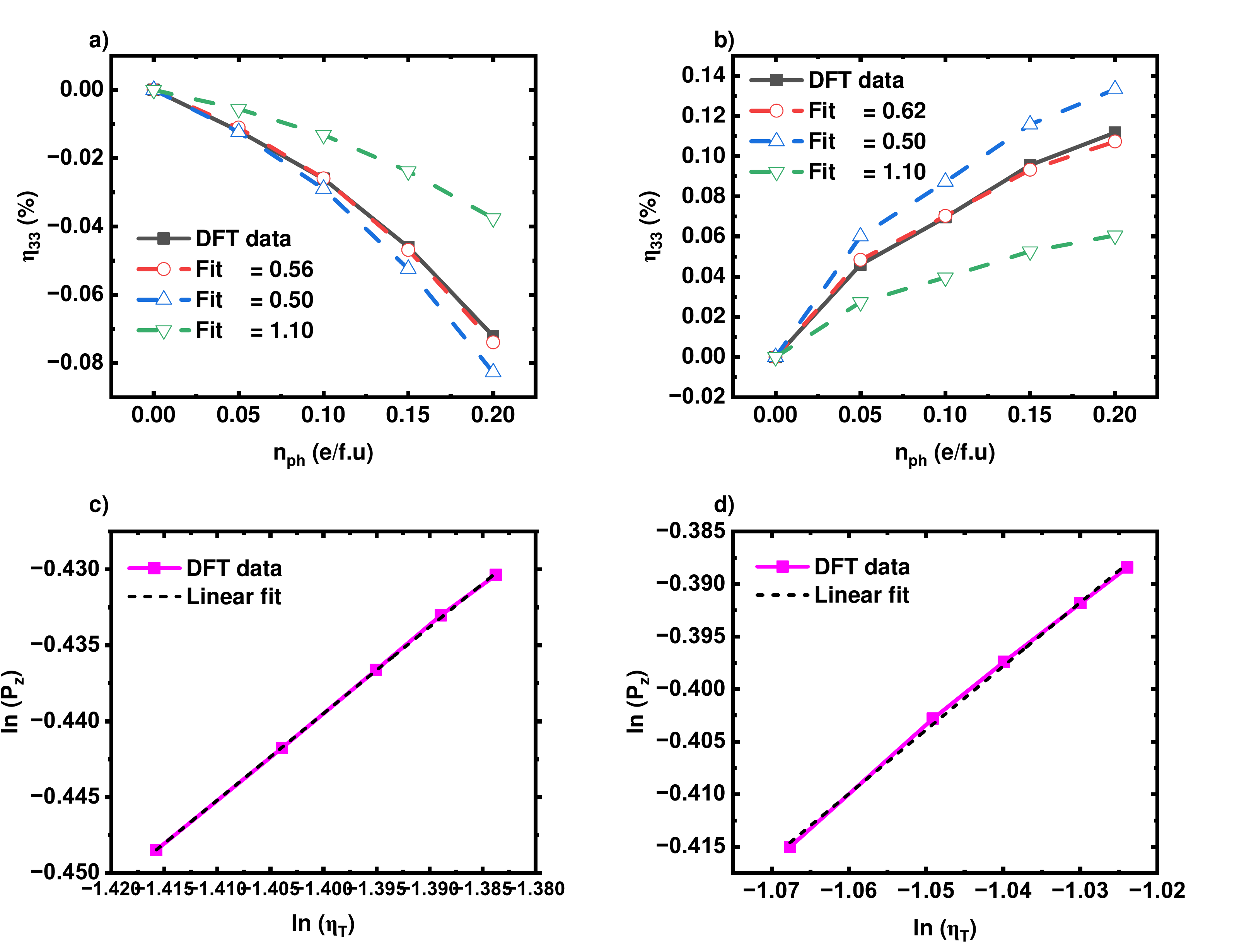}
	\caption{ a), b) plot of the DFT data and fit using Eq. 4 in the main text for the $(2|2)$, $(4|4)$ SLs respectively. c), d) plot of $\ln(P)~ vs ~\ln(\eta_{T})$ for the $(2|2)$, $(4|4)$ SLs respectively.}
\end{figure}
	
\end{document}